\documentclass[onecolumn,floatfix,superscriptaddress,showpacs,showkeys,nofootinbib,preprint]{revtex4}

\usepackage{epsfig}
\usepackage{amssymb,latexsym,amsmath}
\newcommand{\eq}[1]{\begin{align} #1 \end{align}}

\begin{document}


\title{Statistical Ensembles with Volume Fluctuations}

\author{Mark I.~Gorenstein}
\affiliation{Bogolyubov Institute for Theoretical Physics, Kiev, Ukraine}

\begin{abstract}
The volume fluctuations in statistical mechanics are discussed.
First, the volume fluctuations in ensembles with a fixed external
pressure, the so called pressure ensembles, are considered.
Second, a generalization of the pressure ensembles is suggested.
Namely, the statistical ensembles with the volume fluctuating
according to externally given distributions are considered.
Several examples and possible applications in statistical models
of hadron production are discussed.

\end{abstract}



\maketitle
%


\section{Introduction}
Successful application of the statistical model to description of
mean hadron multiplicities in high energy collisions (see, e.g.,
recent papers~\cite{stat1} and references therein) has stimulated
investigations of properties of statistical ensembles.  Whenever
possible, one prefers to use the grand canonical ensemble (GCE)
due to its mathematical convenience. The canonical ensemble
(CE)~\cite{CE} should be applied when the number of carriers of a
conserved charges is small (of the order of 1), such as strange
hadrons~\cite{strange}, anti-baryons~\cite{antibaryons}, or
charmed hadrons \cite{charm}. The micro-canonical ensemble
(MCE)~\cite{MCE} has been used to describe small systems with
fixed energy, e.g. mean hadron multiplicities in proton-antiproton
annihilation at rest. In all these cases, calculations performed
in different statistical ensembles yield different results. This
happens because the systems are `small' and they are `far away'
from the thermodynamic limit (TL). The mean multiplicity of
hadrons in relativistic heavy ion collisions ranges from $10^2$ to
$10^4$, and mean multiplicities (of light hadrons) obtained within
GCE, CE, and MCE approach each other. One refers here to the
thermodynamical equivalence of statistical ensembles and uses the
GCE for calculating the hadron yields.

Measurements of a hadron multiplicity distribution $P(N)$ in
interactions, including nucleus-nucleus collisions, open a new
field of applications of the statistical models. The particle
multiplicity  fluctuations are usually quantified by the ratio of
variance to mean value of a multiplicity distribution $P(N)$, the
scaled variance, and are a subject of current experimental
activities. In statistical models there is a qualitative
difference in the properties of mean multiplicity and scaled
variance of multiplicity distributions. It was recently found
\cite{fluc,mce1,mce2,res,exp,clt,acc} that even in the TL
corresponding results for the scaled variance are different in
different ensembles. Hence the equivalence of ensembles holds for
mean values in the TL, but does not extend to fluctuations.


A statistical system is characterized by the extensive quantities:
volume $V$, energy $E$, and conserved charge(s)\footnote{In
statistical description of the hadron or quark-gluon systems,
these conserved charges are usually the net baryon number,
strangeness, and electric charge. In non-relativistic statistical
mechanics, the number of particles plays the role of a conserved
`charge'.} $Q$. The MCE is defined by the postulate that all
micro-states with given $V$, $E$, and $Q$ have equal probabilities
of being realized. This is the basic postulate of the
statistical mechanics. The MCE partition function just calculates
the number of microscopic states with given fixed $(V,E,Q)$
values.
%
%
%
%
In the CE the energy exchange between the considered system and
`infinite thermal bath' is assumed. Consequently, a new parameter,
temperature $T$ is introduced.
%
%
%
%
To define the GCE, one makes a similar construction for conserved
charge $Q$.  An `infinite chemical bath' and the
chemical potential $\mu$ are introduced.
%
%
%
%
The CE introduces the energy fluctuations. In the GCE, there are
additionally the charge fluctuations.
%
The MCE, CE, and GCE are the most familiar statistical ensembles.
In several textbooks (see, e.g., Ref.~\cite{RR,Tolpygo}), the
pressure (or isobaric) canonical ensemble has been also discussed.
%
%
%
%
The `infinite bath of the fixed external pressure' $p_0$ is then
introduced. This leads to the volume fluctuations around the
average value.

A more general concept of the  statistical ensembles was
suggested in Ref.~\cite{alpha}.  The statistical ensemble is
defined by an externally given distribution of extensive
quantities, $P_{\alpha}(\vec{A})$.  The construction of
distribution of any property $O$ in such an ensemble proceeds in
two steps. Firstly, the MCE $O$-distribution,
$P_{mce}(O;\vec{A})$,  is calculated at fixed values of the
extensive quantities $\vec{A}=(V,E,Q)$. Secondly, this result is
averaged over the external distribution $P_{\alpha}(\vec{A})$
\cite{alpha},
\begin{eqnarray}\label{P}
P_{\alpha} (O) ~=~ \int  d\vec{A}  ~P_{\alpha}(\vec{A}) ~
P_{mce}(O;\vec{A})~.
\end{eqnarray}
The ensemble defined by Eq.~(\ref{P}), the $\alpha$-ensemble,
includes the standard statistical ensembles as particular cases.

Recently, the micro-canonical ensemble with the volume
fluctuations was introduced \cite{MCEsVF} for modelling the hadron
production in high energy interactions. An introduction of the
volume fluctuations was necessary in order to reproduce the
KNO-scaling \cite{KNO} of the hadron multiplicity distribution
observed in proton-proton and proton-antiproton collisions at high
energies. The volume fluctuations lead also to the power law shape
of the single particle spectra at high transverse momenta.

In the present paper  the statistical ensembles with volume
fluctuations are studied.  First,  the pressure ensembles are
considered.  In general, there are 3 pairs\footnote{ In the
present study we do not discuss the role of the total 3-momentum.
As shown in Ref.~\cite{acc} the total momentum conservation is not
important in the TL for thermodynamical functions and fluctuations
in the full phase space. It may however influence the particle
number fluctuations in the limited segments of the phase space.}
of variables -- $(V,p_0),~ (E,T),~ (Q,\mu)$~ -- and, thus, the 8
statistical ensembles\footnote{For several conserved charges
$\{Q_i\}$ the number of possible ensembles is larger, as each
charge can be treated either canonically or grand canonically.}
can be constructed. Among these 8 ensembles there are 4 pressure
ensembles: $(p_0,E,Q)$,~ $(p_0,T,Q)$,~ $(p_0,E,\mu)$,~ and
$(p_0,T,\mu)$. In addition to the pressure canonical ensemble
known from the literature, three other possibilities -- pressure
micro-canonical, pressure grand micro-canonical, and pressure
grand canonical ensembles -- are constructed and studied.
In Section II, non-relativistic statistical
systems
are discussed, whereas  Section III presents the results
for an ultra-relativistic ideal gas.

In Section IV the concept of pressure ensembles is extended to the
case of more general volume fluctuations. Namely, the statistical
ensembles with volume fluctuating according to externally given
distributions are introduced. The Summary presented in Section~V
closes the paper.


\section{Non-Relativistic Boltzmann Gas}
\subsection{Canonical  and Micro-Canonical  Ensembles}
In this Section the system of non-relativistic Boltzmann particles
is discussed. The $(V,T,N)$ {\bf C}anonical {\bf E}nsemble (CE)
partition function of $N$ particles reads \cite{landau}:

\eq{\label{Zce-nr}
Z_{ce}(V,T,N)~=~\frac{1}{N!}\, \int  \frac{d{\bf x}_1d{\bf
p}_1}{(2\pi)^3}~\ldots~\frac{d{\bf x}_Nd{\bf p}_N}{(2\pi)^3}~
      \exp\left[-~\frac{E({\bf x}_1,\ldots,{\bf x}_N;{\bf p}_1,\ldots,{\bf
      p}_N) }{T}\right]~,
%
}
where $T$ is the system temperature, the particle degeneracy
factor is assumed to be equal to 1, and $E$
is the microscopic $N$-particle energy usually presented as the
sum of potential and kinetic terms,
\eq{\label{E-mic}
E({\bf x}_1,\ldots,{\bf x}_N;{\bf p}_1,\ldots,{\bf p}_N)~=~ U({\bf
x}_1,\ldots,{\bf x}_N)~+~\sum_{i=1}^N\frac{{\bf p}_i^2}{2m}~,
}
with $m$ being the particle mass. For the $N$-particle energy
given by~(\ref{E-mic})
the integration over momentum in Eq.~(\ref{Zce-nr}) can be
done explicitly,
\eq{\label{Zce1-nr}
Z_{ce}(V,T,N)~=~\frac{1}{N!}~\left(\frac{mT}{2\pi}\right)^{3N/2}~
%
%
%
%
\int_V  d{\bf x}_1~\ldots~d{\bf x}_N~
      \exp\left[-~\frac{U({\bf x}_1,\ldots,{\bf x}_N)
      }{T}\right]~.
}
The particle coordinate ${\bf x}_1,\ldots,{\bf x}_N$ are
integrated over the system volume $V$. The CE thermodynamical
functions can be expressed in terms of the free energy,
\eq{\label{free}
F(V,T,N)~= -~T~\ln~Z_{ce}(V,T,N)~.
%
}
The CE pressure, entropy, chemical potential, and average energy
are:
\eq{\label{therm-ce}
p~=~-~\left(\frac{\partial F}{\partial V}\right)_{T,N}~, ~~~~S~=~-
\left(\frac{\partial F }{\partial T}\right)_{V,N}~,~~~~
\mu~=~\left(\frac{\partial F}{\partial
N}\right)_{T,V}~,~~~~\overline{E}~=~ F~+~TS~.
 }
For the non-interacting particles, the potential energy  vanishes,
$U({\bf x}_1,\ldots,{\bf x}_N)=0$, thus,
the ideal gas free energy is:
 \eq{\label{free-id}
F_{id}(V,T,N)~
\cong~-NT~-~NT~\ln\left[\frac{V}{N}~\left(\frac{mT}{2\pi}\right)^{3/2}\right]~,
}
where we have assumed $N\gg 1$ and, thus, $\ln N!\cong N\ln N -N$.
The thermodynamical functions of the ideal gas from
Eqs.~(\ref{therm-ce},\ref{free-id}) read:
\eq{\label{pce}
p
=\frac{NT}{V},~~ S=
\frac{5}{2}N+N\ln\left[\frac{V}{N}\left(\frac{mT}{2\pi}\right)^{3/2}\right],
~~\mu=
-T\ln\left[\frac{V}{N}\left(\frac{mT}{2\pi}\right)^{3/2}\right],
%
%
%
%
~~\overline{E}=
\frac{3}{2}NT~.
}

%

\vspace{0.3cm}
The $(V,E,N)$ {\bf M}icro-{\bf C}anonical {\bf E}nsemble (MCE)
partition function of $N$ non-relativistic Boltzmann particles
reads:
\eq{\label{Zmce}
Z_{mce}(V,E,N)~=~\frac{1}{N!}\, \int \frac{d{\bf x}_1d{\bf
p}_1}{(2\pi)^3}~\ldots~\frac{d{\bf x}_N d{\bf p}_N}{(2\pi)^3}~
            \delta\left[E-~U({\bf x}_1,\ldots ,{\bf x}_N)
            -\sum_{j=1}^N~\frac{{\bf p}_j^2}{2m}\right]
%
}
The MCE entropy is defined as,
\eq{\label{Smce}
S(V,E,N)~=~\ln\left[E_0~Z_{mce}(V,E,N)\right]~,
%
}
where $E_0$ is an arbitrary constant with a dimension  of energy.
For  non-interacting particles, $U({\bf x}_1,\ldots,{\bf x}_N)=0$,
one finds,
\eq{\label{Smce1}
Z_{mce}(V,E,N)= \frac{V^N(Em/2\pi)^{3N/2}}{E~N!~\Gamma(3N/2)}~,~~~
S(V,E,N)\cong\frac{5}{2}N
+N\ln\left[\frac{V}{N}\left(\frac{mE}{3N\pi}\right)^{3/2}\right]~.
}
It has been assumed that $N\gg 1$ and $E\gg E_0$, thus,
$\Gamma(3N/2)\cong 3N/2~[\ln(3N/2)-1]$, ~$N!\cong N(\ln N -1)$,
and $\ln(E_0\cdot E^{3N/2-1})\cong \ln(E^{3N/2})$.

The MCE temperature, pressure, and chemical potential are the
following:
\eq{\label{Tmce}
\frac{1}{T}~=~\left(\frac{\partial S}{\partial
E}\right)_{V,N}~
,~~~~~~ \frac{p}{T}~=~\left(\frac{\partial S}{\partial
V}\right)_{E,N}~,~~~~~
%
\frac{\mu}{T}~=~-~\left(\frac{\partial S}{\partial
N}\right)_{V,E}~.
%
}
For non-interacting particles they read,
 \eq{\label{Tmce1}
\frac{1}{T}~=~\frac{3}{2}~\frac{N}{E}~ ,~~~~~~
\frac{p}{T}~=~\frac{N}{V}~,~~~~~
\frac{\mu}{T}~=~-~\ln\left[\frac{V}{N}\left(\frac{mE}{3N\pi}\right)^{3/2}\right]~.
}
If $\overline{E}=E$, the ideal gas system of $N$ particles in the
volume $V$ has the same temperature, pressure, and entropy in the
MCE (\ref{Smce1}-\ref{Tmce1}) and in the CE (\ref{pce}). This
means the thermodynamical equivalence of CE and MCE at $N\gg 1$.

\subsection{Pressure Canonical and Pressure Micro-Canonical Ensembles}
The particle coordinates and momenta for the considered system are
denoted as ${\bf x}_1, ~\ldots, ~{\bf x}_N$ and ${\bf p}_1,
~\ldots, ~{\bf p}_N$, respectively. The $E$ given by
Eq.~(\ref{E-mic}) denotes the system energy, and $V$ is the system
volume. The `thermostat' is now introduced with corresponding
particle coordinates, ${\bf X}_1, ~\ldots, ~{\bf X}_{N_T}$, and
momenta, ${\bf P}_1, ~\ldots, ~{\bf P}_{N_T}$, and with $E_T$ and
$V_T$ being the thermostat energy and volume, respectively. The
system plus thermostat is described by the MCE, i.e. the total
energy and the total volume are assumed to be fixed: $
E+E_T=E^*=$~const~, $V+V_T=V^*=$~const~.
%
The probability distribution of system particle coordinates and
momenta is proportional to,
\eq{\label{fV}
f_V({\bf x}_1, ~\ldots, ~{\bf x}_N; ~{\bf p}_1, ~\ldots, ~{\bf
p}_N)~\propto~\int \prod_{i=1}^{N_T}d^3{\bf X}_i
d^3{\bf P}_i
~\delta\left(E~+~ \sum_{j=1}^{N_T}\frac{{\bf
P}_j^2}{2M}~-~E^*\right)~,
}
where the thermostat particle coordinates ${\bf X}_1,\ldots,{\bf
X}_{N_T}$ are integrated over the thermostat volume $V_T$, and the
system particle coordinates ${\bf x}_1,\ldots,{\bf x}_{N}$ are contained
in the  volume $V$. The particles in the thermostat are assumed to be
non-interacting. Thus, $E_T=\sum_{j=1}^{N_T}{\bf P}_j^2/(2M)$,
with $M$ being the particle mass. The momentum integration in
Eq.~(\ref{fV}) gives,
\eq{\label{p-int}
&\int d^3{\bf P}_1\ldots d^3{\bf P}_{N_T}~\delta\left(E+
\sum_{j=1}^{N_T}\frac{{\bf P}_j^2}{2M}~-~E^*\right)\propto
\left(E^*~-~E\right)^{3N_T/2~-1}~\propto
\left(1-\frac{E}{E^*}\right)^{3N_T/2~-1}~\nonumber
\\
& \cong~\left(1~-~\frac{E}{E_T}\right)^{3N_T/2~-1}
\cong~\left(1~-~\frac{E}{3N_TT/2}\right)^{3N_T/2}~\cong~
\exp(-~E/T)~,
}
where it has been assumed, $N_T\rightarrow\infty$,
$E/E_T\rightarrow 0$, and the ideal gas equation for the
thermostat energy, $E_T=3TN_T/2$, has been used.
The integration over thermostat particle coordinates ${\bf X}_i$
in Eq.~(\ref{fV}) gives the factor depending on the system volume
$V$,
\eq{\label{X-int}
\int_{V_T}~d^3{\bf X}_1\ldots d^3{\bf X}_{N_T}
~&=~V_T^{N_T}~=~(V^*~-~V)^{N_T}~\propto~\left(1~-~\frac{V}{V^*}\right)^{N_T}~
\cong~\left(1~-~\frac{V}{V_T}\right)^{N_T}~\nonumber
\\
&=~\left(1~-~\frac{V}{(TN_T/p_0)}\right)^{N_T}~\cong~\exp
\left(~-~\frac{p_0~V}{T}\right)~,
}
where it has been assumed, $V_T\rightarrow\infty$,
~$N_T\rightarrow\infty$, ~ $V/V_T\rightarrow 0$, and the ideal gas
equation for the thermostat pressure, $p_0=TN_T/V_T$, has been
used.

One obtains from Eqs.~(\ref{fV}-\ref{X-int}),
\eq{\label{f1}
f_V({\bf x}_1, \ldots , {\bf x}_N,~{\bf p}_1, \ldots, {\bf
p}_N)~=~\frac{1}{Z_{pce}(p_0,T,N)}~\frac{1}{N!}~\exp\left(~-~\frac{p_0
V~+~E}{T}\right)~,
}
where the system energy $E$ in (\ref{f1})  depends on particle
coordinates and momenta according to Eq.~(\ref{E-mic}).
The function $Z_{pce}(p_0,T,N)$ in Eq.~(\ref{f1}) is the
$(p_0,T,N)$ {\bf P}ressure {\bf C}anonical {\bf E}nsemble (PCE)
partition function. It is defined by the normalization condition,
\eq{\label{norm}
\int_0^{\infty}dV~\int \prod_{i=1}^{N}\left[\frac{d^3{\bf
x}_id^3{\bf p}_i}{(2\pi)^3}\right] ~f_V({\bf x}_1, \ldots , {\bf
x}_N;~{\bf p}_1, \ldots, {\bf p}_N)~=~1~,
}
and it equals to,
\eq{\label{Zpce}
Z_{pce}(p_0,T,N)~=~\int_0^{\infty}dV~\exp\left(-~\frac{p_0~V}{T}\right)~Z_{ce}(V,T,N)
~,
}
where $Z_{ce}(V,T,N)$ is the CE partition function given by
Eq.~(\ref{Zce-nr}).
%
%
%
%

%
Using $Z_{ce}=\exp(-F/T)$, the probability volume distribution in
the PCE can be presented as:
\eq{\label{V2pce}
W_{pce}(V)~=~
\frac{1}{Z_{pce}(p_0,T,N)}~\exp\left[-~\frac{F(V,T,N)~+~p_0V}{T}\right]~.
}
%
First, one finds the maximum of $W_{pce}(V)$, which defines the most
probable value of the volume, $V=V_0$,
\eq{\label{V0}
\left[\frac{\partial F(V,T,N)}{\partial
V}\right]_{V=V_0}~+~p_0~=~0~.
}
%
By definition, $-~\partial F/\partial V$ is the CE pressure
(\ref{pce}). Thus, the PCE equation of state (\ref{V0}) has clear
physical meaning: the internal pressure $p$ for the most probable
volume $V_0$ equals to the fixed external pressure $p_0$,
\eq{\label{p-p0}
p(V_0,N,T)~=~p_0~.
}
In the TL
the volume distribution (\ref{V2pce}) can be approximated as,
\eq{\label{V1pce}
& W_{pce}(V)~\cong ~\frac{1}{Z_{pce}(p_0,T,N)}~\exp\left[
-~\frac{F(V_0,T,N)~+~p_0V_0}{T}\right]
\\ & \times~\exp\left[-~\frac{1}{2T}\left(\frac{\partial^2 F(V,T,N)}{\partial
V^2}\right)_{V=V_0}\left(V~-~V_0\right)^2\right]
\equiv  \left(2\pi\omega_V~V_0\right)^{-1/2}~
\exp\left[~-~\frac{(V-V_0)^2}{2\omega_V~V_0}\right]~.\nonumber
}
Thus, the most probable volume, $V_0$, and the average volume
$\overline{V}$ are equal to each other in the TL,
\eq{\label{V0-Vav}
\overline{V}~\equiv~\int_0^{\infty}dV~V~W_{pce}(V)~\cong~V_0~,
}
and $\omega_V$ introduced in Eq.~(\ref{V1pce}) defines the scaled
variance of the volume fluctuations,
\eq{\label{omega1V}
\omega_V~\equiv~\frac{\overline{V^2}~-~\overline{V}^2}{\overline{V}}
~=~\frac{T}{V_0}~\left[\frac{\partial^2 F(V,T,N)}{\partial
V^2}\right]_{V=V_0}^{-1}~=~-~\frac{T}{V_0}~\left[\frac{\partial
p(V,T,N)}{\partial V}\right]_{V=V_0}^{-1}~.
}
For the ideal gas pressure (\ref{pce}) one gets:
\eq{\label{id-gas}
\overline{V}~\cong~V_0~=~\frac{NT}{p_0}~,~~~~ \omega_V~\cong~
\frac{T}{p_0}~.
}
The Eqs.~(\ref{V1pce}-\ref{omega1V}) are valid if $(\partial
p/\partial V)$ is negative. This is the case for the `normal'
equation of state, but is not valid
for either a 1st order phase transition or critical point. Note
that the TL $N,V_0\rightarrow\infty$, with $N/V_0$ being finite,
is assumed to make clear notions of the phase transition or
critical point.

\vspace{0.3cm} \noindent{\it 1st Order Phase Transition.} In the
casei of the 1st order phase transition
$\partial p/\partial V=0$ \cite{landau} for
$V_1<V<V_2$ in the mixed phase. According to Eq.~(\ref{V2pce})
this leads to $W_{pce}(V)=const\cong (V_2-V_1)^{-1}$ for
$V_1<V<V_2$. Introducing the notation, $V_2=\gamma V_1$, with
$\gamma= const
>1$, one finds at $V_1\rightarrow\infty$,
\eq{\label{1st}
\overline{V}~&\cong~\int_{V_1}^{V_2}V~W_{pce}(V)dV~=~\frac{\gamma+1}{2}~V_1~,\\
\omega_V&~\cong~
\frac{1}{\overline{V}}~\left[\int_{V_1}^{V_2}V^2~W_{pce}(V)dV~-~\overline{V}^2\right]~
~=~\frac{1}{3}~\left(\frac{\gamma -1}{\gamma
+1}\right)^2~\overline{V} ~.
%
 }
Thus, the volume fluctuations are anomalously large,
$\omega_V\propto \overline{V}\rightarrow\infty$.

\vspace{0.3cm}
 \noindent {\it Critical Point.}   At the critical point it
follows \cite{landau},
\eq{\label{cr-point}
\frac{\partial p}{\partial V}~=~\frac{\partial^2 p}{\partial
V^2}~=~0 ~.
}
In this case, Eq.~(\ref{V2pce}) takes the form,
\eq{\label{V-crit}
&W_{pce}(V)~\cong~\frac{1}{Z_{pce}(p_0,T,N)}~\exp\left[
-~\frac{F(V_0,T,N)~+~p_0V_0}{T}\right]~ \\ &\times
\exp\left[-~\frac{1}{T}~\frac{1}{4!}\left(\frac{\partial^4
F(V,T,N)}{\partial V^4}\right)_{V=V_0}~(V~-~V_0)^4\right]
\equiv~\frac{2~A_0^{1/4}}{\Gamma(1/4)}~\exp\left[~-~A_0~(V~-~V_0)^4\right]~,\nonumber
}
where $A_0\equiv -(24T)^{-1}(\partial^3 p/\partial V^3)_{V=V_0}$.
The volume distribution (\ref{V-crit}) leads to the scaled
variance for the volume fluctuations,
\eq{\label{omega-crit}
\omega_V~=~\frac{\overline{V^2}-\overline{V}^2}{\overline{V}}~\cong~
\frac{1}{V_0}~\int_0^{\infty}dV~(V-V_0)^2~W_{pce}(V)~=
~\frac{1}{V_0}~\frac{\Gamma(3/4)}{\Gamma(1/4)}~A_0^{-1/2}~,
}
where the Gamma functions in Eqs.~(\ref{V-crit}-\ref{omega-crit})
are  $\Gamma(1/4)\cong 3.626$ and $\Gamma(3/4)\cong 1.225$.
Finally, one finds for (\ref{omega-crit}),
\eq{\label{omega1-crit}
\omega_V~\cong~\frac{1.656~T^{1/2}}{V_0}~\left[-~\left(\frac{\partial^3
p}{\partial V^3}\right)_{V=V_0}\right]^{-1/2}~.
}
In order
to estimate the scaled variance (\ref{omega1-crit}) of the volume
fluctuations at the critical point we use the van der Waals (VdW)
equation of state \cite{landau},
\eq{\label{vdw}
\left(p~+~a~\frac{N^2}{V^2}\right)~\left(V~-~b~N\right)~=~NT~.
}
The critical point is defined by the following equations
\eq{\label{p1}
\frac{\partial p}{\partial V}~&
=~-~\frac{NT}{(V-Nb)^2}~+~\frac{2N^2a}{V^3}~=~0~,\\
\frac{\partial^2 p}{\partial V^2}~&
=~\frac{2NT}{(V-Nb)^3}~-~\frac{6N^2a}{V^4}~=~0~.\label{p2}
}
They give,
\eq{\label{cr-vdw}
T_{cr}~=~\frac{8}{27}~\frac{a}{b}~,~~~~V_{cr}~=~3Nb~,~~~~p_{cr}~=~\frac{1}{27}
\frac{a}{b^2}~.
}
 At the critical point one finds,
\eq{\label{p3}
\left(\frac{\partial^3 p}{\partial V^3}\right)_{cr}~
=~\left[-~\frac{6NT}{(V-Nb)^4}~+~\frac{24N^2a}{V^5}\right]_{cr}~=~-~\frac{a}{9^2b^5}~\frac{1}{N^3}~.
}
Thus, for $T=T_{cr}$, $p_0=p_{cr}$, and $\overline{V}\cong
V_{0}=V_{cr}$, and Eqs.~(\ref{omega1-crit},\ref{p3}) give,
\eq{\label{omega2-crit}
\omega_V~\cong~\frac{1.656~T^{1/2}}{V_{cr}}~\left[-~\left(\frac{\partial^3
p}{\partial
V^3}\right)_{V=V_{cr}}\right]^{-1/2}~\cong~2.703~b~\sqrt{N}~.
}
The volume fluctuations at the critical point are anomalously
large, $\omega_V\propto N^{1/2}\propto
\overline{V}^{1/2}\rightarrow\infty$.

\vspace{0.3cm}
Similarly to the PCE we introduce now the $(s_0,E,N)$ {\bf
P}ressure {\bf M}icro-{\bf C}anonical  ensemble (PMC),
\eq{\label{Zpmce}
Z_{pmc}(s_0,E,N)~=~\int_0^{\infty}dV\exp\left(-s_0V\right)Z_{mce}(V,E,N)
~,
}
where the parameter $s_0=p_0/T_0$ is defined by the external
conditions of the thermostat pressure $p_0$ and the thermostat
temperature $T_0$. The micro-canonical energy $E$ is assumed to be
fixed. Thus, there is no energy exchange between the `thermostat'
and considered system, and the internal MCE temperature $T$
(\ref{Tmce}) may differ from $T_0$. Using Eq.~(\ref{Smce}), the
volume distribution in the PMC can be presented in the form,
\eq{\label{Vpmce}
W_{pmc}(V)~&=~\frac{1}{Z_{pmc}(s_0,E,N)}~\exp\left[-~s_0~V~+~S(V,E,N)\right]~
\\ & \cong~
\frac{1}{Z_{pmc}(s_0,E,N)}~\exp\left[-~s_0~V_0~+~S(V_0,E,N)~+~\frac{1}{2}~\left(\frac{\partial
^2 S}{\partial
V^2}\right)_{V=V_0}\left(V-V_0\right)^2\right]~.\nonumber
}
The most probable (and average) volume $V_0$ is defined by the
condition,
\eq{\label{V0-pmce}
-~s_0 ~+ ~\left(\frac{\partial S(V,E,N)}{\partial
V}\right)_{V=V_0}~=~0~.
}
The Eq.~(\ref{V0-pmce}) corresponds to,
\eq{\frac{p}{T}~=~\frac{p_0}{T_0}~,
}
which is similar to Eq.~(\ref{p-p0}) in the PCE. The scaled
variance of the volume fluctuations $\omega_V$ in the PMC equals
to:
\eq{\label{omegaV-pme}
\omega_V~
\cong~-~\left[V_0~(\partial^2 S/\partial
V^2)_{V=V_0}\right]^{-1}~=~-~\frac{T}{V_0}~\left[\frac{\partial
p(V,E,N)}{\partial V}\right]_{V=V_0}^{-1}~,
}
which is again similar to Eq.~(\ref{omega1V}) in the PCE. Thus,
the average volume $V_0$ and scaled variance of the volume
fluctuations $\omega_V$ are identical in the PCE and PMC in the
TL. For the ideal gas  one finds,
\eq{\label{id-pmce}
\langle V \rangle_{pmc}~\cong~
V_0~=~\frac{N}{s_0}~,~~~~~~\omega_V~=~ \frac{1}{s_0}~,
}
which coincide with (\ref{id-gas}) in the PCE.

\subsection{Grand Canonical Ensemble}
The $(V,T,\mu)$ {\bf G}rand {\bf C}anonical {\bf E}nsemble (GCE)
partition function of non-relativistic non-interacting Boltzmann
particles reads:
\eq{\label{Zce}
Z_{gce}(V,T,\mu)~=~\sum_{N=0}^{\infty}~\exp\left(\frac{\mu~N}{T}\right)
~Z_{ce}(V,T,N)~,
%
}
where $Z_{ce}(V,T,N)$ is given by Eq.~(\ref{Zce-nr}) and $\mu$ is
the chemical potential.
%
%
%
%
The GCE pressure is calculated as,
\eq{\label{pTmu}
p(T,\mu)~=~\frac{T}{V}~\ln\left[Z_{gce}(V,T,\mu)\right]~,
}
and it does not depend on the volume $V$ in the TL.  The CE, MCE,
and GCE are thermodynamically equivalent at $V\rightarrow \infty$.
Note that the GCE is the most convenient one from the technical
point of view.

There is, however, an evident problem in the formulation of the
$(p_0,T,\mu)$ {\bf P}ressure {\bf G}rand {\bf C}anonical ensemble
(PGC). The PGC partition function is obtained by extending
Eq.~(\ref{Zpce}),
 \eq{\label{Zpgce-nr}
Z_{pgc}(p_0,T,\mu)~=~
\int_0^{\infty}dV~\exp\left(-~\frac{p_0~V}{T}\right)~Z_{gce}(V,T,\mu)~
=~\frac{T}{p_0~-~p(T,\mu)}~,
}
where $p(T,\mu)$ is the GCE pressure (\ref{pTmu}).  The
$(p_0,T,\mu)$-ensemble has a unique property. Among
8 possible ensembles this is the only one where the system
description includes only intensive quantites,
$p_0,T$ and $\mu$.
For
$p(T,\mu)=p_0$, the system volume is undefined.
In the domain $p(T,\mu)\ge p_0$, the PGC partition function
does not exist as the integral over the volume
in Eq.~(\ref{Zpgce-nr})
diverges. For $p(T,\mu)<p_0$, the volume distribution has the
form:
 \eq{\label{WVpgce-nr}
W_{pgc}(V)~=~ \frac{1}{Z_{pgc}(p_0,T,\mu)}~
\exp\left[-~V~\left(\frac{p_0~-~p}{T}\right)\right]~.
}
%
The most probable volume $V_0$ equals to zero, but the average
volume is:
 \eq{\label{Vpgce-nr}
\overline{V}~=~
\int_0^{\infty}dV~V~W_{pgc}(V)~
=~\frac{T}{p_0~-~p(T,\mu)}~.
}
%
This special ensemble will be discussed further in details for  the
ultra-relativistic gas in the next section.

\section{Ultra-Relativistic Gas}
In this section several examples of the pressure ensembles for the
ultra-relativistic ($m=0$) ideal gas of Boltzmann particles are
considered.
For simplicity only
statistical systems without conserved
charges are discussed. Thus, the number of particles is not
restricted and chemical potential equals to zero.
Furthermore,
the Boltzmann statistics is used and the degeneracy factor is
assumed to be one.

\subsection{Grand Canonical and Grand Micro-Canonical Ensembles}
The $(V,T)$ GCE\footnote{The chemical potential connected to the
number of particles equals to zero.} partition function of
massless non-interacting neutral Boltzmann particles reads:
\eq{\label{Zgce}
Z_{gce}(V,T)=\sum_{N=0}^{\infty}\frac{1}{N!}
       \left(\frac{
       V}{2\pi^2}\right)^N\int_0^{\infty}\prod_{i=1}^{N}p^2_idp_i
\exp\left(-\frac{p_i}{T}\right)
 =\sum_{N=0}^{\infty}\frac{1}{N!}~\left(\frac{VT^3}{\pi^2}\right)^N=
 \exp(\overline{N})~,
%
}
where $\overline{N}\equiv \langle N\rangle_{gce} =VT^3/\pi^2$ is
the GCE average number of particles.
The GCE system pressure
and average energy are:
\eq{\label{gce-p-rel}
p~&=~\frac{T}{V}~\ln Z_{gce}~=~\frac{T^4}{\pi^2}~=~T~n(T)~,\\
\langle E\rangle_{gce}~&\equiv~\overline{E}~=~T^2~\frac{\partial
\ln Z_{gce}}{\partial T}~=~
\frac{3}{\pi^2}~VT^4~=~\varepsilon(T)~V~,\label{gce-E-rel}
%
}
where $n(T)=\overline{N}/V=T^3/\pi^2$ and
$\varepsilon(T)=3T^4/\pi^2$ are the particle number density and
energy density, respectively. The GCE multiplicity distribution
has the Poisson form,
\eq{\label{PN-gce}
P_{gce}(N;V,T)~=~\frac{\overline{N}^N}{N!}~\exp\left(-~\overline{N}\right)~,
}
and the scaled variance of particle number distribution
(\ref{PN-gce}) equals to:
\eq{\label{omega-gce}
\omega_{gce}~\equiv~ \frac{\langle N^2\rangle_{gce} ~-~\langle
N\rangle^2_{gce}}{\langle N \rangle_{gce}}~=~1~.
}

%

 \vspace{0.3cm}
The $(V,E)$ {\bf G}rand {\bf M}icro-{\bf C}anonical
ensemble\footnote{The energy is fixed, but the number of particles
is not. The chemical potential connected to the number of
particles equals to zero. Thus, this ensemble is named the Grand
Micro-Canonical Ensemble \cite{mce2}.} (GMC) partition function is
\cite{mce1}:
\begin{align}\label{WE}
Z_{gmc}(V,E)
\;&=~\sum_{N=1}^{\infty}\frac{1}{N!}\,\left(\frac{V}{2\pi^2}
\right)^N~\int_0^{\infty}\prod_{i=1}^{N}p_i^2dp_i~\delta\left(E~-~\sum_{j=1}^N
p_j\right) \;\nonumber
       \\
~&\equiv \; \sum_{N=1}^{\infty}W_N(V,E)
      \;=\; \frac{1}{E}\sum_{N=1}^{\infty}\frac{A^N}{N!~(3N-1)!}~=~\frac{A}{2E}\;\;_0F_3
     \left(;\,\frac{4}{3},\frac{5}{3},2;\,\frac{A}{27}\right)~,
\end{align}
where $_0F_3$ is the generalized hyper-geometric function
\cite{I},
%
and
\begin{align}\label{xdef}
A~ \equiv ~\frac{VE^3}{\pi^2}~.
\end{align}
%
The GMC particle number distribution function equals to:
\begin{align}\label{PNmce}
P_{gmc}(N;V,E)~\equiv~\frac{W_{N}(V,E)}{Z_{gmc}(V,E)} \;=\;
\frac{1}{Z_{gmc}(V,E)}~\frac{A^N}{E~N!~(3N-1)!}\;.
\end{align}
It is defined for $N\ge 1$.
%
%
%
The average number of particles in the GMC equals to \cite{mce1}:
\begin{align}\label{Ngce1}
\langle N\rangle_{gmc} ~\cong~ \left(\frac{A}{27}\right)^{1/4}\;.
\end{align}
In the large volume limit the mean multiplicities in the GME and the
GCE are equal, $\langle N\rangle_{gmc} = \overline{N}$
providing $E=\overline{E}$ and the GMC and the GCE
volumes are equal.
The temperature and the
pressure in the GME are equal to:
\eq{\label{T-gme}
T~=~\frac{E}{3\overline{N}}~=~\left(\frac{\pi^2E}{3V}\right)^{1/4},~~~~~~p
~=~\frac{\overline{N}T}{V}~=~\frac{T^4}{\pi^2}~,
}
and they coincide with the corresponding quantities
(\ref{gce-p-rel}-\ref{gce-E-rel}) in the GCE.
%

For $\overline{N}\gg1$
the particle number distribution in the GME (\ref{PNmce}) can be
approximated by the Gaussian one:
\begin{align}\label{gaussN}
P_{gmc}(N;V,E) \;\cong ~\left(2\pi~\omega_{gce}~
\overline{N}\right)^{-1/2}~
 \exp\left[~-~\frac{\left(N - \overline{N}\right)^2}
 {2~ \omega_{gmc}~  \overline{N}}\right]~,
\end{align}
with the GME scaled variance
\eq{
\omega_{gmc}~=~\frac{\langle N^2\rangle_{gmc}~-~\langle
N\rangle_{gmc}^2}{\langle N\rangle_{gmc}}~=~\frac{1}{4}~.
}
The Poisson distribution $P_{gce}(N)$ (\ref{PN-gce}) for
$\overline{N}\gg 1$ can be also approximated by the Gauss
distribution (\ref{gaussN}), but with $\omega_{gce}=1$.
%
%
%
%
In the TL the particle number distributions in the GME and GCE
have both the Gauss form \cite{clt}. The average number of
particles is the same, $\langle N\rangle_{gmc}\cong \overline{N}$,
but the scaled variance is different,
$\omega_{gce}=1$ and $\omega_{gmc}=1/4$.

%
\subsection{Pressure Grand Canonical Ensemble}

The $(p_0,T)$ {\bf P}ressure {\bf G}rand {\bf C}anonical ensemble
(PGC) is defined as the ensemble with the fixed external pressure
$p_0$, temperature $T$, and the chemical potential connected to
the number of particles  equals to zero. The PGC partition
function equal to \cite{pressure}:
\eq{\label{Zpgce}
Z_{pgc}(p_0,T)~
\equiv~\int_0^{\infty}dV~\exp\left(-~\frac{p_0V}{T}\right)~Z_{gce}(V,T)~=~
\frac{T}{p_0~-~p(T)}~,
 }
where the relation (\ref{gce-p-rel}) between $p(T)$ and $Z_{gce}$
has been used. The value of $p_0$ has a physical meaning of the
external pressure. A convergence of the integral over the volume
in Eq.~(\ref{Zpgce}) requires the inequality,
\eq{\label{T01}
p(T)~=~\frac{1}{\pi^2}~T^4~<~p_0~.
}
Thus, at each fixed value of $p_0$, there is a `limiting
temperature' $T^*$ in the PGC:
\eq{\label{T0}
T~<~T^*~=~\left(\pi^2~p_0\right)^{1/4}~.
}
%

%

The probability volume distribution in the PGC is:
\eq{\label{W-pgce}
W_{pgc}(V)~=~\frac{1}{Z_{pgc}(p_0,T)}~\exp\left[-~V\left(\frac{p_0~-~p(T)}{T}\right)\right]~.
}
It gives the most probable volume equal to zero, $V_0=0$,
and the average value,
\eq{\label{V-pgce}
\langle V
\rangle_{pgc}~\equiv~\overline{V}~=~\int_0^{\infty}dV~V~W_{pgc}(V)~=~-~T~\frac{\partial
\ln Z_{pgc}(p_0,T)}{\partial p_0}~=~\frac{T}{p_0~-~p(T)}~,
}
which is not equal to zero and may even go to infinity at $T
\rightarrow T^*$. Using Eq.~(\ref{V-pgce}), the volume
distribution (\ref{W-pgce}) can be written as,
\eq{\label{W-pgce1}
W_{pgc}(V)~=~\overline{V}^{~-1}~\exp\left(-~V/\overline{V}\right)~.
}
For the mean values of energy and particle number  in the
PGC one finds:
\eq{
 \langle E \rangle_{pgc}~&=~
\int_0^{\infty}dV~W_{pgc}(V)~\langle
E\rangle_{gce}~=~\frac{3}{\pi^2}~T^4~\overline{V}~=~
\varepsilon(T)~\overline{V},\label{E-pgce}\\
\langle N \rangle_{pgc}~&=~ \int_0^{\infty}dV~W_{pgc}(V)~\langle
N\rangle_{gce}~=~
\frac{1}{\pi^2}~T^3~\overline{V}~=~n(T)~\overline{V}\label{N-pgce}~,
}
where $n(T)$ and $\varepsilon (T)$ are the particle number density
and energy density of the GCE given by Eq.~(\ref{gce-p-rel}) and
Eq.~(\ref{gce-E-rel}), respectively.

The independent variables are $(V,T)$ in the GCE and $(p_0,T)$ in
the PGC.
If the GCE volume is chosen to be equal to the average volume of
the PGC, then Eqs.~(\ref{E-pgce}) and (\ref{N-pgce}) give:
$\langle E \rangle_{pgc}=\overline{E}$ and $\langle N
\rangle_{pgc}=\overline{N}$,
i.e. the GCE and PGC are thermodynamically equivalent. The
fluctuations of $E$ and $N$ are however  different in these
two ensembles.
Calculating,
\eq{
\langle V^2
\rangle_{pgc}~=~\int_0^{\infty}dV~V^2~W_{pgc}(V)~=~\frac{T^2}{Z_{pgc}}~
\frac{\partial^2 Z_{pgc}}{\partial
p_0^2}~=~2~\left[\frac{T}{p_0~-~p(T)}\right]^2~=~2\overline{V}^2~,
}
one finds,
\eq{\label{omega-Vpgce}
\omega_V~=~\frac{\langle V^2 \rangle_{pgc}~-~\langle V
\rangle_{pgc}^2}{\langle V \rangle_{pgc}}~=~\overline{V}~.
%
}
Thus, in the TL limit $\overline{V}\rightarrow\infty$ the volume
fluctuations become anomalously large. This, in turn, leads to
anomalous energy and particle number fluctuations. The particle
number distribution in the PGC has the form of the geometrical
distribution:
\eq{\label{PN-pgce}
P_{pgc}(N;p_0,T)~=~ \int_0^{\infty}dV~W_{pgc}(V)~P_{gce}(N,V,T)
~=~(1-\eta)~\eta^N~,
}
where $P_{gce}(N,V,T)$ has been taken from Eq.~(\ref{PN-gce}), and
$\eta \equiv (T/T^*)^4<1$~.  The most probable number of particles
is $N=0$, whereas the average value (\ref{N-pgce}) is larger than zero.
It
equals to $\langle N \rangle_{pgc}=\eta (1-\eta)^{-1}$ and may
even go to infinity at $\eta\rightarrow 1$. This happens if
$T\rightarrow T^*$. From Eq.~(\ref{PN-pgce}) it follows:
%
%
%
%
%
\eq{\label{omega-pgce}
\omega_{pgc}~=~\frac{\langle N^2 \rangle_{pgc}~-~\langle N
\rangle_{pgc}^2}{\langle N
\rangle_{pgc}}~=~1~+~\frac{\eta}{1-\eta}~=~1~+ n(T)~\overline{V}~.
}
%
%
The first term in the r.h.s. of Eq.~(\ref{omega-pgce}) corresponds
to the Poisson fluctuations (\ref{omega-gce}) of the GCE at fixed
volume, whereas the second term comes from the volume fluctuations
at fixed particle number density.
%
%
%
%
The multiplicity distribution (\ref{PN-pgce}) can be rewritten as,
\eq{\label{PN-pgce1}
P_{pgc}(N;p_0,T)~\equiv~P_{pgc}(N;\overline{N})~=~\frac{1}{\overline{N}+1}~\exp\left[-~N~
\ln\left(1+\frac{1}{\overline{N}}\right)\right]~.
}
For $\overline{N}\gg 1$ the distribution $P_{pgc}$ approches:
\eq{\label{PN-pgce2}
P_{pgc}(N;\overline{N})~\cong~\frac{1}{\overline{N}}~\exp\left(-~\frac{N}{\overline{N}}\right)~.
}
The particle number distribution $P_{pgc}(N)$ (\ref{PN-pgce2})
satisfies the so called KNO-scaling \cite{KNO}.


\subsection{Pressure Grand Micro-Canonical Ensemble}

The $(s_0,E)$ {\bf P}ressure {\bf G}rand {\bf M}icro-canonical
ensemble (PGM) is defined as the statistical ensemble with fixed
energy $E$ and fixed external parameter $s_0=p_0/T_0$. As before,
the chemical potential connected to the number of particles equals
to zero. The PGM partition function is equal to:
\eq{\label{Z-pmce}
&Z_{pgm}(s_0,E)~\equiv~\int_0^{\infty}dV~\exp(-~s_0V)~Z_{gmc}(V,E)~ \\
&=~
\frac{1}{E}~\sum_{N=1}^{\infty}\left(\frac{E^3}{\pi^2}\right)^N~
\frac{1}{N!~(3N-1)!}\int_0^{\infty}dV~\exp(-~s_0V)~V^N
~=~\frac{1}{s_0~E}~\sum_{N=1}^{\infty}\frac{B^N}{(3N-1)!}~,\nonumber
}
where
\eq{\label{y}
B~\equiv~\frac{E^3}{s_0~\pi^2}~.
%
}
Using Eq.~\cite{Prud},
%
 \eq{\label{PBM}
\sum_{k=0}^{\infty}~\frac{x^{3k}}{(3k)!}~=~\frac{1}{3}\left[\exp(x)~+~2\exp(-x/2)
\cos \left(\frac{\sqrt{3}}{2}~x\right)\right]~,
}
%
%
one finds at $B\gg 1$,
\eq{\label{Z1-pmce}
\ln[Z_{pgm}(s_0,E)]~\cong~B^{1/3}~.
}

In the TL, the volume distribution in the PGM can be found from
(\ref{Z-pmce}) using the asymptotic behavior of $_0F_3$ function
\cite{I},
\eq{
\ln[Z_{gmc}(V,E)] ~=~\ln\left[\frac{A}{2E}\;\;_0F_3
     \left(;\,\frac{4}{3},\frac{5}{3},2;\,\frac{A}{27}\right)\right]~\cong~
4~\left(\frac{VE^3}{27\pi^2}\right)^{1/4}~.
}
The volume distribution in the PGM is then proportional to,
\eq{
W_{pgm}(V)~&\propto~\exp\left[4\left(\frac{VE^3}{27\pi^2}\right)^{1/4}~-~s_0V\right]
~\equiv~\exp[\phi(V)]~\nonumber\\
&\cong~\exp\left[\phi(V_0)~+~\frac{1}{2}\left(\frac{\partial^2\phi}{\partial
V^2}\right)_{V=V_0}~\left(V~-~V_0\right)^2 \right]~.
}
The most probable volume $V_0$ in the PGM is defined by the
condition,
\eq{
\left(\frac{\partial \phi}{\partial
V}\right)_{V=V_0}~=~\left(\frac{E^3}{27\pi^2V_0^3}\right)^{1/4}~-~s_0~=~0~.
}
This gives,
\eq{\label{V0-pgm}
V_0~=~\frac{E}{3\pi^{2/3}s_0^{4/3}}~.
}
One also finds,
\eq{
\left(\frac{\partial^2\phi}{\partial
V^2}\right)_{V=V_0}~=~-~\frac{3s_0}{4V_0}~.
}
Thus, the volume distribution in the PGM can be approximated as:
\eq{\label{W-pgm}
W_{pgm}(V)~\cong~\left(2\pi~\omega_V~V_0\right)^{-1/2}~\exp\left[-~\frac{\left(V~-~V_0\right)^2}
{2~\omega_V~V_0}\right]~,
}
where $\omega_V=4/(3s_0)$ is the scaled variance of the volume
fluctuations in the PGM.  The average volume and its fluctuations
in the PGM can be also calculated in terms of the partition
function $Z_{pgm}(s_0,E)$ using Eqs.~(\ref{Z-pmce}-\ref{Z1-pmce}),
\eq{\label{V1-pmce}
\langle
V\rangle_{pgm}~&=~\int_0^{\infty}dVW_{pgm}(V)~V~=~-~\frac{\partial
\ln ~Z_{pgm}}{\partial s_0} ~\cong~\frac{E}{3\pi^{2/3}s_0^{4/3}}~,\\
\langle
V^2\rangle_{pgm}~&=~\int_0^{\infty}dVW_{pgm}(V)~V^2~=~\frac{1}{Z_{pgm}}~
\frac{\partial^2 Z_{pgm}}{\partial
s_0^2}~\cong~
\frac{E^2}{9\pi^{4/3}s_0^{8/3}}~,\label{V2-pmce} \\
\omega_V~&=~\frac{\langle V^2\rangle_{pgm}~-~\langle
V\rangle^2_{pgm}}{\langle
V\rangle_{pgm}}~\cong~\frac{4}{3s_0}~.\label{omegaV-rel}
}
The condition (\ref{V0-pgm}) can be written as,
\eq{\label{p-s}
s_0~=~\frac{1}{\pi^2}~T^3~=~\frac{p}{T}~,
}
and it means that the internal pressure $p$ equals to $s_0T$.
The most probable volume $V_0$ (\ref{V0-pgm}) and average volume
$\langle V\rangle _{pgm}$ (\ref{V1-pmce}) are then equal to each
other, and both are equal to $E\pi^2/(3T^4)$. This corresponds to
the fixed volume in the MCE with fixed energy $E$ and the MCE
temperature $T$. The volume distribution in the PGM is therefore
different from that  in the PGC (\ref{W-pgce1}). In contrast
to the PGC, there is an extensive variable, the energy $E$, in the
PMG. This leads to the system average volume proportional to the
energy and given by Eq.~(\ref{V0-pgm}). The internal pressure
equals to $p=s_0T$ (\ref{p-s}).  As a result, the volume
fluctuations in the PGM are
Gaussian~(\ref{W-pgm})  with
the finite scaled variance~(\ref{omegaV-rel}).
Using~Eq.~(\ref{p-s}) the scaled variance (\ref{omegaV-rel}) of
the volume fluctuations can be expressed in terms of the MCE
temperature, $\omega_V=4\pi^2/(3T^3)$.
One also finds:
\eq{\label{Nrel-pmce}
\langle N\rangle_{pgm}~&=~\frac{1}{Z_{pgm}}~B~\frac{\partial
Z_{pgm}}{\partial B}~\cong~\frac{1}{3}~B^{1/3}~,\\
\langle
N^2\rangle_{pgm}~&=~\frac{1}{Z_{pgm}}~B\frac{\partial}{\partial B}
B~\frac{\partial Z_{pgm}}{\partial B}~\cong~ \frac{1}{9}~B^{1/3}~
+~\frac{1}{9}~B^{2/3},\\
\omega_{pgm}~&=~\frac{\langle N^2\rangle_{pgm}~-~\langle
N\rangle^2_{pgm}}{\langle
N\rangle_{pgm}}~\cong~\frac{1}{3}~.\label{omegarel-pmce}
}
In the TL, the multiplicity distribution in the PGM can be
approximated as:
\begin{align}\label{gaussN2}
P_{pgm}(N;s_0,E) \;\cong ~\left(2\pi
   ~\omega_{pgm}~ \overline{N}\right)^{-1/2}~
 \exp\left[~-~\frac{\left(N - \overline{N}\right)^2}
 {2~ \omega_{pgm}~  \overline{N}}\right]~,
\end{align}
where $\overline{N}=B^{1/3}/3\cong \langle N\rangle_{pgm}$ and
$\omega_{pgm}=1/3$~.
%
%
%
%
%
The scaled variance in the PGM can be presented in the TL as the
following,
\eq{\label{omegarel1-pmce}
\omega_{pgm}~=~\omega_{mce}~+~\frac{1}{16}~\omega_V~n~=~\frac{1}{4}~+~\frac{1}{12}
~=~\frac{1}{3},
}
where $n=T^3/\pi^2$ is the MCE particle number density. The first
term in the r.h.s. of Eq.~(\ref{omegarel1-pmce}),
$\omega_{mce}=1/4$~, is due to the particle number fluctuations in
the MCE with fixed volume, and the second term is the contribution
due to the volume fluctuations.


\section{Ensembles with External Volume Fluctuations}
The multiplicity distributions $P(N)$ in relativistic gases
\cite{fluc,mce1,mce2,res,exp,clt,acc} are sensitive to
conservation laws obeyed by the system, and therefore to
fluctuations of extensive quantities $E$ and $Q$. The examples
considered in the previous section demonstrate that the volume
fluctuations also influence the particle number fluctuations.
Thus, for the calculation of multiplicity distributions, the
choice of the statistical ensemble is then not a matter of
convenience, but a physical question. On the other hand, the
fluctuations of extensive quantities $\vec{A}\equiv (V,E,Q)$
depend not on the system's physical properties, but rather on
external conditions. One can imagine a huge variety of these
conditions, thus, 8 statistical ensembles discussed  in the
previous sections are only some special examples.

A more general concept of the statistical ensembles based on
Eq.~(\ref{P}) was recently suggested in Ref.~\cite{alpha}. The
system volume may exhibit fluctuations described by the externally
given distribution. When $V$ is the only fluctuating variable,
Eq.~(\ref{P}) is reduced  to
\eq{\label{PalphaN1a}
P_{\alpha} (O) = \int dV ~P_{\alpha}(E,V,Q) ~  P_{mce}(O;E,V,Q)~,
}
where $P_{\alpha}(V)$ is externally given volume distribution.

The effect of volume fluctuations on the particle
number fluctuations is calculated for
the system of non-interacting massless
Boltzmann particles with zero chemical potential. At fixed volume
the system is treated within the GMC, and the particle number
distribution is $P_{gmc}(N;V,E)$ (\ref{PNmce}).

In the {\it first example}, the volume distribution is assumed to be:
\eq{\label{P-a-V}
P_{\alpha}(V)~=~\left(2\pi\omega_V~a_V^2\overline{V}\right)^{-1/2}~
 \exp \left[- ~ \frac{\left( V~-~\overline{V}
    \right)^2}{2\omega_V~a_V^2\overline{V}} \right]~,
}
where $\omega_V = 4\pi^2/(3T^3)$ coincides with (\ref{omegaV-rel})
in the PGM.
The choice of $P_{\alpha}(V)$ results in a simple correspondence
to the GMC and PGM in the TL. In Eq.~(\ref{P-a-V}), $a_V$ is a
dimensionless tuneable parameter which determines the width of the
distribution. In the limit $a_V \rightarrow 0$, Eq.~(\ref{P-a-V})
becomes a Dirac $\delta$-function, $\delta(V-\overline{V})$. This
corresponds to the GMC.  For $a_V = 1$, Eq.~(\ref{P-a-V}) results
in the PGM volume fluctuations (\ref{W-pgm}) in the TL.
The particle number distribution reads,
\begin{eqnarray}\label{P-a-N}
P_{\alpha} (N) ~=~ \int_0^{\infty}  dV ~P_{\alpha}(V) ~
P_{gmc}(N;V,E)~.
\end{eqnarray}
A substitution of $P_{\alpha}(V)$ in Eq.~(\ref{P-a-N}) by the
distribution (\ref{P-a-V}) results in $P_{\alpha}(N)$ in the TL
given by the Gaussian,
\eq{
 \label{p-N-aV}
  P_{\alpha} (N)
~ \cong~ \left(2 \pi \omega_{\alpha}\overline{N}\right)^{-1/2}
~\exp \left[ - ~\frac{\left( N~-~\overline {N} \right)^2}{ 2
\omega_{\alpha}~\overline{N}}\right]~,
}
where the average number of particles $\overline{N}$ is defined by
the energy and average volume, $\overline{N}=~
[\overline{V}~E^3/(27\pi^2)]^{1/4}$, and the scaled variance of
the particle number distribution (\ref{p-N-aV}) equals,
\eq{\label{omega-aV}
\omega_{\alpha}~=~\frac{1}{4}~+~\frac{1}{12}~a_V^2~.
}
The first term in the r.h.s. of Eq.~(\ref{omega-aV}) equals to the GMC
scaled variance at fixed $E$ and $V$, the second term is due to
the volume fluctuations. As it can be expected,
$\omega_{\alpha}=\omega_{gmc}$ for $a_V=0$, and $\omega_{\alpha} =
\omega_{pgm}$ for $a_V=1$. It also follows,
$\omega_{gmc}<\omega_{\alpha}<\omega_{pgm}$ for $0<a_V<1$, and
$\omega_{\alpha}>\omega_{pgm}$ for $a_V>1$. The $\alpha$-ensemble
defined by Eqs.~(\ref{P-a-N},~\ref{P-a-V}) presents an extension of
the GMC ($a_V =0$) and PGM ($a_V =1$) to a more general volume
distribution.

\vspace{0.3cm} In the {\it second example}   both $E$
and $V$ are assumed to fluctuate.  Equation~(\ref{P-a-N})
should be then extended as:
\eq{\label{PalphaN1}
P_{\alpha} (N) ~=~ \int dEdV ~P_{\alpha}(E,V) ~  P_{gmc}(N;E,V)~.
}
First, uncorrelated volume and energy
distributions are considered:
\eq{\label{PEV}
P_{\alpha}(E,V)~=~P_{1}(V)\times P_{2}(E)~,
}
where $P_{1}(V)$ is given by Eq.~(\ref{P-a-V}) and $P_{2}(E)$ is
taken in the following form,
\eq{\label{P-a-E}
P_{2}(E)~=~\left(2\pi\omega_E~a_E^2\overline{E}\right)^{-1/2}~
 \exp \left[- ~ \frac{\left( E~-~\overline{E}
    \right)^2}{2\omega_E~a_E^2~\overline{E}} \right]~,
}
where $\overline{E}=3T^4\overline{V}/\pi^2$ and $\omega_E =4T$. A
substitution of $P_{\alpha}(E,V)$ in Eq.~(\ref{PalphaN1}) by the
above distributions results in $P_{\alpha}(N)$ in the TL given
again by the Gaussian (\ref{p-N-aV})
%
%
%
%
with the average number of particles $\overline{N}$ defined by the
average energy and average volume, $\overline{N}=~
[\overline{V}~\overline{E}^3/(27\pi^2)]^{1/4}$, and the scaled
variance equal to:
\eq{\label{omegaEV}
\omega_{\alpha}~=~\frac{1}{4}~+~\frac{3}{4}~a_E^2~+~\frac{1}{12}a_V^2~.
}
The first term in the r.h.s. of Eq.~(\ref{omegaEV}) is the MCE
scaled variance at fixed $E$ and $V$, the second term is due to
energy fluctuations, and the third one comes from volume
fluctuations. The results of the GMC, PGM, and GCE for the scaled
variance of the particle number distribution can be obtained from
Eq.~(\ref{omegaEV}) at $a_V=a_E=0$; $a_V=1, ~a_E=0$; and $a_V=0,
~a_E=1$, respectively

Examples of correlated $V$ and $E$ distributions can be
constructed as follows. One assumes the factorized form
(\ref{PEV}) with Eq.~(\ref{P-a-E}) for the energy distribution,
but with $\overline{E}=3T^4V/\pi^2$ and $\omega_E =4T$, i.e. the
average energy $\overline{E}$ depends on the volume $V$. Assuming,
%
 \eq{\label{P-1-V}
P_{1}(V)~=~\frac{1}{\overline{V}}~\exp(-~V/\overline{V})~,
%
}
with $\overline{V}$ given by Eq.~(\ref{V-pgce}), from
Eqs.~(\ref{PEV}) and (\ref{P-a-E}) with $a_E=1$,  the
results for the PGC are obtained.

The relation, $\overline{E}=3T^4V/\pi^2$, can be used
together with any form of the
volume distribution $P_1(V)$.
This yields a generalization of the GCE to the
systems with externally given volume fluctuations. The energy
fluctuations at fixed $V$ can be also selected in accordance
with the physics requirements.

As the {\it third example} of the statistical ensembles with
volume fluctuations one refers to the recent paper \cite{MCEsVF}
where the micro-canonical ensemble with the scaling volume
fluctuations for the ultra-relativistic ideal gas has been
considered. The volume fluctuations were assumed to have the
specific scaling properties. They were chosen to describe the KNO
scaling \cite{KNO} of the particle multiplicity distributions
measured in proton-proton collisions at high energies. A striking
feature of the model is power law form of the single momentum
spectrum at high momenta, instead of the exponential Boltzmann
distribution in the systems with fixed volume.

\section{Summary}
In this paper the volume fluctuations in the statistical mechanics
have been studied. The statistical ensembles with fixed external
pressure were considered. Statistical systems of classical
non-relativistic particles and non-interacting massless particles
were discussed. The volume fluctuations in the pressure canonical
ensemble become anomalously large in the thermodynamic limit for
the first order phase transition, $\omega_V\propto N$, or at the
critical point, $\omega \propto N^{1/2}$. Another type of
anomalous volume fluctuations takes place for the pressure grand
canonical ensemble. In this special  ensemble, all thermodynamical
variables are the intensive quantities, $(p_0,T,\mu)$. In the
thermodynamical limit the mean particle multiplicity obtained
within the considered ensembles with volume fluctuations is equal
to the mean calculated within the micro-canonical, canonical, and
grand canonical ensembles. This is not valid, however, for the
scaled variance. The influence of the volume fluctuations in the
pressure ensembles on particle number fluctuations have been
discussed for ultra-relativistic ideal gas.

Following Ref.~\cite{alpha} the ensembles with volume fluctuations
determined by the externally given distribution function were
introduced. Multiplicity fluctuations are sensitive to the volume
fluctuations. The volume fluctuations may also influence the
behavior of the particle momentum spectra \cite{MCEsVF}. Thus, we
believe that the concept of statistical ensembles with fluctuating
extensive quantities, in particular, with fluctuating volume of
the statistical system, may be appropriate for the statistical
description of hadron production in relativistic collisions. It
may be also useful for other physical systems. In fact, in all
cases when the equilibrium statistical mechanics is used to
calculate the fluctuations of the system properties.

\begin{acknowledgments}
We would like to thank D.V.~Anchishkin, V.V.~Begun,
M.~Ga\'zdzicki, W.~Greiner, M.~Hauer, B.I.~Lev, I.N.~Mishustin,
O.N.~Moroz, and Yu.M.~Sinyukov for numerous discussions. This work
was in part supported by the Program of Fundamental Researches of
the Department of Physics and Astronomy of NAS, Ukraine.
\end{acknowledgments}


\end{document}